# Sample Size Considerations in the Design of Orthopaedic Risk-factor Studies

Richard Evans [†] and Antonio Pozzi [*]

† Corresponding Author

† Clinical and Translational Science Institute
University of Minnesota
Minneapolis, Minnesota, USA

* Small Animal Clinic
University of Zurich
Zurich, Switzerland

No funding was received for this research, and there are no conflicts of interest.

Richard Evans contributed to the concept, writing, and simulations in this manuscript.
Antonio Pozzi contributed to the concept and writing of this manuscript.

# Abstract

## Objective

Sample size calculations play a central role in study design because sample size affects study interpretability, costs, hospital resources, and staff time. For most orthopaedic risk-factor studies, either the sample size calculation or the post-hoc power calculation assumes the disease status of control subjects is perfectly ascertained, when it may not be. That means control groups may be mixtures of both unaffected cases and some unidentified affected cases. Negative control groups with misclassified positive data are called unlabeled. Treating unlabeled groups as disease-negative control groups is known to cause misclassification bias, but there has been little research on how misclassification affects the statistical power of risk association tests. In this study, we demonstrate the consequences of using unlabeled groups as control groups on the power of risk association tests, with the intent of showing that control groups with even small misclassification rates can reduce the power of association tests. So, sample size calculations that ignore the unlabeled aspect of control groups may underpower studies. In addition, we offer a range of correction factors to adjust sample size calculations back to 80% power.

## Study Design

This was a simulation study using study designs from published orthopaedic risk-factor studies. The approach was to use their designs but simulate the data to include known proportions of misclassified affected subjects in the control group. The simulated data was used to calculate the power of a risk-association test. We calculated powers for several study designs and misclassification rates and compared them to a reference model.

## Results

Treating unlabeled data as disease-negative only always reduced statistical power compared to the reference power, and power loss increased with increasing misclassification rate. For this study, power could be improved back to 80% by increasing the sample size by a factor of 1.1 to 1.4.

## Conclusion

Researchers should use caution in calculating sample sizes for risk-factor studies and consider adjustments for estimated misclassification rates.

# Introduction

Sample size calculations play a central role in study design because sample size affects study costs, hospital resources, and staff time. Also, underpowered studies with Type II errors can be hard to interpret and there are also serious ethical considerations for studies that have little chance of success..

In orthopaedic risk factor studies, the positive disease status of the affected subjects is ascertained with perfect sensitivity and specificity, but sometimes the disease status of control subjects is not perfectly ascertained. That means control groups may be mixtures of both unaffected cases and some unidentified affected cases. Control groups with misclassified data are called unlabeled. Data with truly affected cases in the positive group and an unlabeled control group is called positive-unlabeled (PU) data by the data science community.

Examples of positive-unlabeled data are well documented in human medicine, but less so in veterinary medicine. Nevertheless, many veterinary studies fall into the positive-unlabeled framework. For example, genome-wide association studies of cranial cruciate ligament disease (CCLD) in dogs use case-control designs. The affected cases are truly positive CCLD cases because they are enrolled from the set of dogs who have undergone knee stabilization surgery. The control cases are typically five years old or older with no history of CCLD and pass an orthopaedic veterinary exam by a board-certified surgeon. However some control dogs will have spontaneous rupture in the future, and so genetically belong in the CCLD affected group. Other control dogs may have sub-diagnostic disease. For example, a dog might appear sound on physical exam and be enrolled in the control group, but may actually have force-platform-detectable hindlimb lameness. Such a dog should not be in the control group because the lameness might be subclinical CCLD.

There are other examples of PU data in the veterinary literature, typically in risk-factor studies using case-control designs. For example, Arthur et al. (2016) used a case-control design to assess the risk of osteosarcoma following fracture repair. They said, "There may be additional cases [in the control group] in which implant-related osteosarcoma was diagnosed in the private practice setting without referral…," suggesting that the control group may be unlabeled because some control-group cases were actually osteosarcoma positive, but diagnosed outside the study. In another example, Wylie et al. 2013 studied risk factors for equine laminitis using controls obtained from an owner survey. The authors noted the positive-unlabeled aspect of their data, "Our study relied on owner-reported diagnoses of endocrinopathic conditions, and this may have introduced misclassification bias."

As mentioned above, the affected cases are "labeled" positive, but the control data is "unlabeled," because dogs may be affected or unaffected. Treating the unlabeled control group as entirely unaffected is called the naive model. The proportion of affected dogs in the control group is called the nondetection rate or undetected rate.

Using the naive model when the nondetection rate is positive causes misclassification bias (because there are affected cases in the control group), and that bias is well documented in the data science literature. Biases due to misclassification can be mitigated using models other than the naive model and with the appropriate data analysis, and there are many articles describing methods for analyzing positive-unlabeled data. Bekker and Davis (2020) provide an excellent summary of methods. Sometimes, however, researchers prefer the naive model because the analysis is simpler and they believe their small nondetection rates induce misclassification biases that are too small for practical consideration. There is some suggestion that nondetection rates under 10% do have little impact on bias.

But bias in estimates (e.g., bias in regression coefficients) is just one part of the results; the other part is inference (e.g., p-values). Central to inference is the power of statistical tests. Power is used in planning a study as a measure of the ability of the study to make the correct decisions. That is, finding P<0.05 when it should. Typically, 80 percent power means that if the group parameters are truly different, then the statistical test has an 80 percent chance of obtaining p<0.05.

During the design phase of risk association studies, researchers might calculate the sample size they need for 80% power assuming the nondetection rate is zero. That is, there are no misclassified affected subjects in the control group. However, if after collection the data are positively unlabeled, then the naive model is incorrect and the estimated power may be less than estimated. We investigated the effect of positive-unlabeled data on loss of statistical power under the naive model. For comparison, the reference power is defined as the power when the naive model is correct and the group sizes are balanced. The results are described in terms of power loss relative to the reference power, both percent power loss and absolute power loss. For context, these two quantities are analogous to relative risk and absolute risk from epidemiology.

Using a simulation, we described how statistical power changes with varying proportions of undetected positives in the naive controls, and varying the imbalance between the numbers of cases and naive controls. Our first aim was to demonstrate that the naive analysis of positive-unlabeled data reduces statistical power in risk-factor studies, even for small nondetection rates. Our second aim was to offer correction factors to upward-adjust sample sizes and correct for the power loss described in Aim 1.

## Methods and materials

### The Test of Association

This was a simulation study assessing the changes in the power of a univariate association test under different PU conditions. There are many statistical tests of association, but we calculated the power for one of the most common tests, Fisher's exact test, which is used to test the statistical significance of a binary risk factor. More generally, this test can be used to assess the significance of any risk factor using the predicted values from a univariate logistic regression.

In the context of risk association studies, and all else being equal, Fisher's exact test would achieve its maximum power for a balanced study design when the naive model is correct

(i.e., no undetected positives in the control group). We call that maximum power the reference power and reported our results as both percent power loss relative to the reference power and as absolute power loss from reference power. In other words, we are using Fisher's exact test to show how much statistical power might be lost by ignoring the nondetection rate.

## The Sample Size and Group Imbalance

The total sample size for the simulation was fixed at N=200, which is consistent with Healey et al. 2019 (N=216), and Baird et al. 2014 (N=217). The effect size, 0.21, was chosen because, with N=200, the reference power was close to 80 percent, which is a value that is commonly used in study design. That way, the reference model is the one with a standard power of 80 percent. Note that the sample size and effect size are not key parameters for the simulation because for any sample size an effect size can be chosen so that power is 80 percent. Also, effect size and sample size are not features of PU data, per se.

The simulation study varied two study design parameters: the nondetection rate and group-size imbalance. The proportion of undetected positives in the control group ranged from 0 (the value for reference power) to 10 percent. We used 10 percent as the upper limit because researchers are generally willing to accept nondetection rates below 10 percent and use the naive model, but change to a PU analysis for rates greater than 10 percent.

We modeled group imbalance using Healey et al. (2019), which used 161 dogs affected with CCLD and 55 unlabeled dogs as controls, and Baird et al. (2014) which used 91 dogs affected with CCLD, and 126 unlabeled dogs as controls, so that imbalance ratios were about 3:1 and 1:3. We only used two imbalance proportions (1:3 and 3:1) and no imbalance (1:1) because the key parameter for this study was the nondetection proportion. That gave simulation sample sizes of (50, 150), (150, 50), and (100, 100).

## The Simulation Algorithm

The overall approach is to simulate data, and then use that data to calculate the p-value of Fisher's exact test. The process is repeated 5000 times for each combination of sample size and nondetection rate, and then the 5000 p-values are compared to 0.05. The proportion of p-values less than 0.05 is the estimated power.

Simulating the data works backward from what might be expected. Instead of starting with values for a risk factor (e.g., 200 0's and 1's representing sex) and then simulating their disease status, we start with the disease status (e.g., 50 affected cases and 150 controls with 135 unaffected and 15 affected) and then assign binary values for the risk factor. It was done that way to control the nondetection rate and group sizes.

The simulation algorithm is most easily described using examples, and we begin with calculating power for Fisher's exact test under the reference model, which is 100 cases and 100 correctly labeled (i.e., 100 truly unaffected) controls. That is, there are no affected cases in the control group, so this is not positive-unlabeled data, and the naive model is the correct model. Next, we associate a binary risk factor variable, $X$ (e.g., sex), with the cases and controls. The negative controls were simulated by sampling 100 negative cases from a binomial distribution with $Pr(X = 1) = 0.2$. That probability means the the baseline risk for the disease in the population is 0.2. It it was chosen arbitrarily, because the baseline risk isn't central to power, the effect size is. As mentioned above, the effect size was 0.21, so the 100 positive cases were sampled from a binomial distribution with $Pr(X = 1) = 0.2 + 0.21$. Using the sex example, that means that having sex=1 predisposes the animals to about double the baseline risk of disease (the baseline is 0.2, and with sex = 1, the risk is double, 0.2 + 0.21 = 0.41.).

Now, the 200 cases are pairs of binary data, one representing the group and the other representing the risk factor. These simulated data were tested with Fisher's exact test. As mentioned above, this process was repeated 5000 times, and the resulting 5000 p-values used to estimate power.

For the second example, we calculate the power for a positive-unlabeled example. Suppose that in a 100-patient control group, 10 percent are in fact undetected positives. So the dataset is 10 affected cases in the control group, 90 unaffected cases in the control group, and 100 affected cases in the positive group. As in the previous example, the risk factor is simulated by sampling from binomial distributions. Now, 90 controls are sampled from the binomial distribution with $Pr(X = 1) = 0.2$, the 10 affected controls are sampled from the binomial distribution with $Pr(X = 1) = 0.2 + 0.21$, and 100 cases are sampled from the same binomial distribution with $Pr(X = 1) = 0.2 + 0.21$. The 10 mislabeled affected cases remain in the control group, so as to measure the effect of treating PU data naively. As before, the simulated data were treated like pilot data, and p-values were calculated. This process is repeated 5000 times and the was estimated as described in the previous example.

### The Correction Factor

For aim 2, the sample-size correction factor estimation, we used the same simulation algorithm and effect size (0.21) but multiplied the group sample sizes by possible correction factors, 1.1, 1.2, and so on, increasing sample size and therefore the power, until it reached the 80%.

### Results

Table 1 describes power loss for the three study designs with three different group sizes, (50, 150), (150, 50), and (100, 100), and for three nondetection rates, 0, 0.05 (5%), and 0.1 (10%). To give these parameters context, if the group sizes are (50, 150) and the nondetection rate is 0.1, then the positive (affected) group has 50 cases, and the unlabeled control group (which we are treating naively in the analysis) has 150 cases, 15 of which are actually affected cases. When the nondetection rate is zero, the naive model is correct because there are no affected cases in the control group. The first row is the reference power, so its loss of power compared to itself is zero. The reference power was calculated in the simulation just like all the other powers and was estimated to be 0.82.

Columns 5 and 6 are the power loss columns and have negative entries because for this simulation positive-unlabeled data analyzed under the naive model always had lower power, as did unbalanced data. Column 5 is the percent loss from the reference power (82%) and column 6 is the absolute power loss from the reference power. For example, the second row shows a -4.81% relative power reduction when the group sizes are balanced but five percent (0.05) of the control group are actually positive cases.

Rows one, four, and seven are correct models (i.e., no positives in the control group). Rows four and seven show a power loss due to sample size imbalance only. So, for this small example, group imbalance sometimes caused more power loss than misclassified data as is seen by comparing row 3 to row four. It is known that for equal overall sample size, group imbalance results in less powerful tests. As an aside, more data is often better than less data, and it is sometimes better to have more unbalanced data than fewer balanced data.

Using Table 1, increasing the nondetection rate within a study design decreased power. For example, for the (100, 100) study design, power decreased by more than 10% as the non-detection rate increased (rows one to three). For the (50, 150) design, in rows seven to nine, power decreased by 12.02% (22.47 - 10.45) from the correct model (row seven), but 22.47% from the reference model. Finally, note that for this simulation, the absolute power losses are marked, but not extreme. For example, in the (100, 100) design, (rows 1 to 3) the power dropped to 0.73 (0.82 - 0.09) for row 3.

Table 1. Power loss. This table orders sample sizes by relative power loss (%). The first row is the reference power, which had an absolute power of 0.82 (82%). The last two columns represent power loss relative to 0.82, both as a percentage and an absolute difference. Note that some inconsistencies in the table are due to rounding. For example, in rows 3 to 5, the absolute power is constant at 0.09, but the relative power changes.

| Row | N positive cases | N naive controls | nondetection proportion | Relative power loss (%) | Absolute power loss (from 0.82) |
|---|---|---|---|---|---|
| 1 | 100 | 100 | 0.00 | 0.00 | 0.00 |
| 2 | 100 | 100 | 0.05 | -4.81 | -0.04 |
| 3 | 100 | 100 | 0.10 | -10.29 | -0.09 |
| 4 | 150 | 50 | 0.00 | -10.77 | -0.09 |
| 5 | 150 | 50 | 0.05 | -14.92 | -0.13 |
| 6 | 150 | 50 | 0.10 | -22.50 | -0.20 |
| 7 | 50 | 150 | 0.00 | -10.45 | -0.09 |
| 8 | 50 | 150 | 0.05 | -15.26 | -0.13 |
| 9 | 50 | 150 | 0.10 | -22.47 | -0.20 |

Table 2 shows how many additional subjects are required to regain power when the non-detection rate is 10%. Rows 1 to 5 are for the (100, 100) design, rows 6 to 10 are for the (150, 50) design, and rows 11 to 15 are for the (50, 150) design. The sample size was increased by 10% for each row within a study design. As one might expect, lower positive-unlabeled power needs more subjects to bring the power up to 80%. For the unbalanced designs, the increased sample size also fixes the power loss due to imbalance. For example, in row one, the power is 0.69 for the original (50, 150) design with 10% nondetection rate. Row 14 shows that an additional 65 subjects, or 32.5% more subjects are required to bring the power above 80%. However, for the (100, 100) design, only 10% more subjects are required (rows one and two).

Table 2. Power improvement. Rows 1, 6, and 11, show the power for when there are no false positives. The other rows show improvements in power when there is a 10% nondetection rate. The sixth column shows the percent increase in sample size, and the last column is power.

| Row | N positive cases | N naive controls | N false controls | N total | percent increase in N | Power |
|---|---|---|---|---|---|---|
| 1 | 100 | 100 | 10 | 200 | 0.0 | 0.78 |
| 2 | 110 | 110 | 11 | 220 | 10.0 | 0.83 |
| 3 | 120 | 120 | 12 | 240 | 20.0 | 0.87 |
| 4 | 130 | 130 | 13 | 260 | 30.0 | 0.89 |
| 5 | 140 | 140 | 14 | 280 | 40.0 | 0.90 |
| 6 | 150 | 50 | 5 | 200 | 0.0 | 0.67 |
| 7 | 165 | 55 | 6 | 220 | 10.0 | 0.70 |
| 8 | 180 | 60 | 6 | 240 | 20.0 | 0.77 |
| 9 | 195 | 79 | 8 | 274 | 37.0 | 0.84 |
| 10 | 210 | 80 | 8 | 290 | 45.0 | 0.87 |
| 11 | 50 | 150 | 15 | 200 | 0.0 | 0.69 |
| 12 | 55 | 165 | 16 | 220 | 10.0 | 0.73 |
| 13 | 60 | 180 | 18 | 240 | 20.0 | 0.78 |
| 14 | 70 | 195 | 20 | 265 | 32.5 | 0.83 |
| 15 | 80 | 210 | 21 | 290 | 45.0 | 0.85 |

## Discussion

This study showed that under specific conditions there was modest power loss even for relatively small proportions of undetected positives in the control group. That means that risk-factor studies may have lower than expected power, and therefore increased chance of Type II error. It is important to note that changes in power may affect more than power. For example, in Table 1, rows 1 to 3, absolute power dropped from 0.82 (row 1) to 0.73 (row 3). It would take about 20 additional subjects to reach the reference power (using Table 2). If subjects were very expensive, then the apparently small drop in power (0.09) is actually large in terms of cost. On the other hand, for an exploratory retrospective study, a 0.09 (9%) power drop may not be considered much.

The working examples were from GWAS studies, but the simulation results apply to any kind of study with univariate association tests, such as any risk factor study. That is a broad class of studies. Examples include the univariate associations between post-op surgical infections and various surgical conditions (e.g., boarded surgeon vs. resident, manufacturer of bone plates). In that case, there may be subdiagnostic infections in the control group. Another example is univariate association tests in veterinary surveys.

In this simulation, the undetected positives in the negative group were randomly sampled from the same population as the detected positives in the affected group. That's a common assumption, but there are other models. In one such model, the undetected positives in the control group might be a subpopulation of positives defined by another variable. For example, in a CCLD GWAS study, the undetected positives in the control group might be positive dogs with low body condition scores. We did not explore those kinds of models in this research. Our goal was to find some examples to show that for some studies, misclassified data may cause power loss. When that power loss is combined with unbalanced data, the loss can be extreme.